\documentclass[universe,article,accept,moreauthors,pdftex]{Definitions/mdpi}

\makeatletter
\g@addto@macro{\UrlBreaks}{\UrlOrds}
\makeatother

\usepackage{booktabs}

\usepackage[utf8]{inputenc}
\usepackage{amsmath}
\usepackage{amssymb}
\firstpage{1} 
\pubvolume{6}
\issuenum{8}
\articlenumber{108}
\pubyear{2020}
\copyrightyear{2020}
\history{Received: 23 June 2020; Accepted: 31 July 2020; Published: date}
\updates{yes} % If there is an update available, un-comment this line

\Title{Revisiting the Cosmological Constant Problem within~Quantum Cosmology}
\Author{Vesselin G. Gueorguiev $^{1,2,*}$%\hl{*}
\orcidA{} and Andre Maeder $^{3}$\orcidB{}}

\AuthorNames{Vesselin G. Gueorguiev and Andre Maeder}

\address{%
$^{1}$ \quad Institute for Advanced Physical Studies, Montevideo Street, 1618 Sofia, Bulgaria\\
$^{2}$ \quad Ronin Institute for Independent Scholarship, Montclair, NJ 07043  USA\\
$^{3}$ \quad Geneva Observatory, University of Geneva, 
51, chemin des Maillettes, CH-1290 Sauverny, Switzerland; 
Andre.Maeder at unige.ch
}

\corres{Correspondence: Vesselin at MailAPS.org}

\abstract{
A new perspective on the Cosmological Constant Problem (CCP) is proposed and discussed 
within the multiverse approach of Quantum Cosmology. 
It is assumed that each member of the ensemble of universes has a characteristic scale $a$ 
that can be used as integration variable in the partition function. 
An averaged characteristic scale of the ensemble is estimated 
by using only members that satisfy the Einstein field equations. 
The averaged characteristic scale is compatible with the Planck length 
when considering an ensemble of solutions to the Einstein field equations 
with an effective cosmological constant. 
The multiverse ensemble is split in Planck-seed universes with 
vacuum energy density of order one; thus, $ \tilde{\Lambda}\approx 8\pi$ 
in Planck units and $a$-derivable universes. 
For~$a$-derivable universe with a characteristic scale of the order of 
the observed Universe $a\approx 8\times10^{60}$, the cosmological constant 
$\Lambda=\tilde{\Lambda}/a^{2}$ is in the range $10^{-121}$--$10^{-122}$, 
which is close in magnitude to the observed value $10^{-123}$.
We point out that the smallness of $\Lambda$ can be viewed to be natural 
if its value is associated with the entropy of the Universe.
This approach to the CCP reconciles the Planck-scale huge vacuum energy--density 
predicted by QFT considerations, as valid for Planck-seed universes, 
with the observed small value of the cosmological constant as relevant to 
an $a$-derivable universe as~observed.
}

\begin{document}
\maketitle

\nolinenumbers %VGG
%\tableofcontents
%MDPI: It is better to delete contents part, please confirm.
%VGG:Yes, it is ok!
 
\section{Introduction}

Constructing a theory that unites successfully the ideas behind general
relativity and quantum mechanics is expected to yield a new understanding
of physical reality. Currently, there are conceptual and technical
problems that one is forced to face when thinking of such fusion.
However, progress in observational cosmology seems to provide new
puzzles along with the necessary guidance for the development and
verification of the relevant physical models \cite{Spergel2007,Komatsu2011,PlanckCollaboration2016}.

Understanding the small positive value of the cosmological constant
$\Lambda$ and the associated energy density $\rho_{\Lambda}$ is
one of the biggest problems in modern physics, qualitatively and quantitatively.
Currently, the Quantum Field Theory (QFT) estimate of $\rho_{\Lambda}$
is about $120$ orders of magnitude bigger than the actually measured
value \cite{Hawking1983,Weinberg1989,Spergel2007,Komatsu2011,PlanckCollaboration2016}.
There is a big variety of models that are attempting to explain the
value of $\rho_{\Lambda}$. Most models consider dynamical mechanisms
to generate $\rho_{\Lambda}$ as zero-point energy (vacuum energy)
of a matter field. In the 1980s, there were few very appealing general
arguments that $\Lambda$ should be zero or can be viewed as an integral
of the motion, some arguments were based on quantum gravity~\cite{Coleman1988},
some on supersymmetry \cite{Hawking1983}, or reparametrization invariance \cite{Weinberg1989}. 
However, experimental data have shown that $\rho_{\Lambda}$ is actually small and positive: 
\begin{equation}
\rho_{\Lambda}=\rho_{c}\Omega_{\varLambda}=\frac{c^{4}}{8\pi G}\Lambda\sim6.3\times10^{-9}~\text{erg/cm}^{3},\label{eq:rho_Lambda_exp}
\end{equation}
which is $1.4\times10^{-123}$ times smaller than the Planck's scale
energy density: 
\begin{equation}
\rho_{Planck}=\frac{c^{7}}{\hbar G^{2}}\sim4.6\times10^{114}~\text{erg/cm}^{3}.\label{eq:rho_Planck}
\end{equation}

Here, $\rho_{c}=3H_{0}^{2}c^{2}/(8\pi G)$ is the critical energy--density,
$H_{0}=70~\text{km\:s}^{-1}\text{Mpc}^{-1}$ is the Hubble expansion rate, 
and $\Omega_{\varLambda}=0.72$ based on observations 
 \cite{Komatsu2011,Hinshaw2013,PlanckCollaboration2016}.
The definition of the critical energy--density is a naturally important
quantity---it separates the open from the closed cosmological solutions
within the cosmology based on Friedmann–Lemaitre–Robertson–Walker (FLRW) metric.

Seeing the enormous discrepancy above is just the numerical manifestation
of a fundamental problem in our understanding of nature. Thus, looking
at different viewpoints and lines of reasoning can be enlightening.
Let's briefly sketch a few of them: of course, one shall start with
the source: Einstein and his ``biggest blunder''. As the legend
has it, Einstein introduced the cosmological constant $\Lambda$ into
his equation $G_{\alpha\beta}=\kappa T_{\alpha\beta}$ in order to
counteract the gravitational pull of the matter encoded by $T_{\alpha\beta}$---the stress--energy--momentum tensor. $T_{\alpha\beta}$ determines
the geometry of the space-time via the metric tensor $g_{\alpha\beta}$,
which is a solution to the Einstein equation,
and the associated metric connection $\nabla_{\alpha}$ that provides
the covariant derivative of various tensors. In Einstein's GR, the
metric tensor $g_{\alpha\beta}$ and the connection are in special
relation $\nabla_{\gamma}g_{\alpha\beta}=g_{\alpha\beta;\gamma}=0$
and are utilized to build the Einstein tensor $G_{\alpha\beta}=R_{\alpha\beta}-\frac{1}{2}g_{\alpha\beta}R$
with the help of the scalar curvature $R=R_{\alpha\beta}g^{\beta\alpha}$
and the Ricci tensor $R_{\alpha\beta}=R_{\:\alpha\gamma\beta}^{\gamma}$
constructed from the Riemann curvature tensor $R_{\alpha\beta\gamma\nu}$.

The Einstein's equation $G_{\alpha\beta}=\kappa T_{\alpha\beta}$
relies on the matter conservation laws $\nabla_{\gamma}T^{\alpha\gamma}=0$
and the uniqueness of the Einstein tensor with the same property 
$\nabla_{\gamma}G^{\alpha\gamma}=0$;
then, the constant of proportionality $\kappa=\frac{8\pi G}{c^{4}}$
is determined from the Newtonian limit of the theory. According to
this reasoning, the Einstein's equations are uniquely determined and
illustrate the idea that the matter distribution and motion determine
the metric tensor $g_{\alpha\beta}$ and thus the structure of the
space-time. In particular, when there is no matter and only vacuum
$T^{\alpha\gamma}=0$, the solution corresponds to Ricci flat spacetime
since $G_{\alpha\beta}g^{\beta\alpha}=-R=0$ and thus 
$R_{\alpha\beta}=G_{\alpha\beta}=0$.
However, if there is matter, then,~depending on the initial conditions,
the solutions to the Einstein's GR equations will correspond to systems
that re-collapse or expand forever. Thus, the desire for a static
solution prompted the introduction of a term $\Lambda g_{\alpha\beta}$
into the Einstein's GR equation 
$G_{\alpha\beta}+\Lambda g_{\alpha\beta}=\kappa T_{\alpha\beta}$,
where $\Lambda$ is a constant since the new term has to obey the
conservation laws already satisfied by $G_{\alpha\beta}$ and $T_{\alpha\beta}$
along with the postulate $g_{\alpha\beta;\gamma}=0$. When $\Lambda$
is introduced in the model, as illustrated by the above line of reasoning,
the result is a phenomenological theory with a constant parameter
$\Lambda$ that has to be determined by experiment and observations.
Furthermore, if one is to estimate $\Lambda$ by using dimensional
reasoning based on the relevant system parameters such as $c$, $G$,
and the size of the Universe via the use of the Hubble constant $H_{0}$,
then one obtains an energy density (\ref{eq:rho_H0}) that is very
close to the experimental value above (\ref{eq:rho_Lambda_exp}):
\begin{equation}
\rho_{H}\propto\frac{2H_{0}^{2}}{\kappa c^{2}}=
\frac{1}{4\pi} \frac{H_{0}^{2}c^{2}}{G}
=6\times10^{-9}~\text{erg/cm}^{3}\label{eq:rho_H0}
\end{equation}

Here, the numerical factor of $4\pi$ can be justified as the 
solid angle being part of the volume measure for the energy density in 3D space.
Therefore, by inspection of (\ref{eq:rho_Lambda_exp}) and (\ref{eq:rho_H0}),
one can conclude that $\Lambda \approx 2H_{0}^{2}/c^{2}$ and the
Hubble constant $H_{0}$ should be one of the fundamental constants of nature. 
It~is important that the expression above is based on simple dimensional reasoning. 
However, one~can get a very similar result
($\Lambda\approx 3H_{0}^{2}/c^{2}$) by considering the LFRW equations 
that are derived from the Einstein field equations with cosmological constant 
by using the Robertson-Walker metric ansatz 
and neglecting the curvature and matter energy--density terms. 
Note that such approximations are standard handwaving estimates in building
simple models of nature. The Robertson--Walker metric is the most general
homogeneous and isotropic metric (up to time-dependent conformal factor), 
thus~consistent with the cosmological principle that the Universe is 
homogeneous and isotropic at very large scales. 
One can neglect the curvature and matter
energy--density terms since there seems to be very little matter in
the Universe at very large scales and the Universe seems to be large
enough to appear practically flat. In this respect, the cosmological constant 
is a property of an expanding, flat, and matter empty Universe, 
which except for its  expanding property is what one would call a classical~vacuum.

An alternative point of view is to reason that, even if one neglects
the contribution of the matter and electromagnetic radiation to the
stress--energy--momentum tensor $T_{\alpha\beta}$, thus setting $T_{\alpha\beta}=0$
as for classical \textit{matter vacuum}, there is still energy and
momentum within the space-time due to the possible presence of gravitational
waves. Therefore, there should be a non-zero source on the right-hand
side of the Einstein equation $G_{\alpha\beta}=-\Lambda g_{\alpha\beta}$
\footnote{The negative sign is for consistency of moving $\varLambda$ from
the RHS to the LHS of the equation, as well as for convenience in
relating $\varLambda$ to the scalar curvature $R$.}
that should be related to the presence of a metric field $g_{\alpha\beta}$.
This~way, the parameter $\Lambda$ of proportionality has something
to do with stress--energy--momentum tensor for gravity in space-time
of the classical matter vacuum. Solution to such equation has been introduced
by de Sitter and has illustrated that matter may not be the only source
of gravity and thus wrinkles in the space-time may not be due to matter only. 
Furthermore, such solutions illustrate that vacuum solutions do not have to be Ricci
flat but instead can correspond to non-zero constant scalar curvature
$R=4\Lambda$. This way one extends minimally the Einstein's GR framework
in the sense that space-time can be spatially flat, homogeneous, and isotropic, 
but with a non-zero constant scalar curvature. The possibility of non-flat 
universe has recently received observational support \cite{Valentino2020} .
Thus, $\Lambda$ can be viewed as a 
fundamental constant which is somehow related to the energy--density of the vacuum. 
Furthermore, since the Einstein original equation, $G_{\alpha\beta}=\kappa T_{\alpha\beta}$, 
now reads as $G_{\alpha\beta}=-\Lambda g_{\alpha\beta}$, therefore, the stress--energy--momentum 
tensor can be viewed as a tensor that contains (non-classical) non-zero vacuum part:
\begin{equation}
\kappa T_{\alpha\beta}=-\Lambda g_{\alpha\beta}.
\label{eq:Tand_Lambda}
\end{equation}

Due to the signature of the metric $g_{\alpha\beta}$, such vacuum has a very special equation of state, 
relationship between pressure and energy--density, $p_{\rm vac}= -\rho_{\rm vac}$ 
that behaves thermodynamically well upon adiabatic expansion and compression. 
This point of view implies that $\rho_{\rm vac}$ is constant. 
Thus, $\rho_{\rm vac}$ is related to the cosmological constant in a very simple way 
$\Lambda=\kappa \rho_{\rm vac}$ \cite{Carroll1992}.
Unfortunately, the scale of the energy--density that one can deduce based on 
the relevant fundamental constants like $c$, $G$, and the Planck constant $\hbar$, 
results than in $\rho_{Planck}$ (\ref{eq:rho_Planck}), which is about $120$ orders of magnitude
bigger than the actually measured value of $\rho_{\Lambda}$ (\ref{eq:rho_Lambda_exp}).

The above discussion demonstrates that ``old fashioned'' dimensional
reasoning provides a reasonable value for $\Lambda$ and the corresponding
energy--density $\rho_{\Lambda}$, while the quantum field theory based
reasoning results in an enormously different energy--density value. 
The smallness of the observed cosmological constant and the related energy--density
is one of three puzzling facts. If the cosmological constant is really due to quantum effects,
then the other puzzle is why quantum energy fluctuations are not manifesting at
the scale of applicability of the general relativity. The third question is related 
to the comparable value of $\rho_{\Lambda}$ to the overall matter density 
$\rho_{m}$, which is about a factor of three given that $\Omega_m=28\%$.
%MDPI: Remove all blue color from main text, please confirm.
%VGG:Yes, it is ok!
It seems that the  last two questions could be addressed well within the 
unimodular gravity approach to  Einstein's equations \cite{Smolin2009}. 
While the smallness of $\Lambda$ could be justified in few possibly equivalent ways:
via the Universe as a Black Hole idea where the large entropy in the Universe
results is small $\Lambda$ (\ref{eq:SdS}), or via the quantum corrected 
Raychaudhuri equation \cite{Farag&Ali&Das2015}, 
or as a second-order quantum correction to the Newtonian gravity 
\cite{Chiarelli2019, Chiarelli2020}, or as related to the 
relevant observables at the infrared stable fixed point of the 
4D gravity at large distances \cite{Antoniadis&Mottola1992}.
It may not be a surprise that all these could be equivalent, 
but, to the best of our knowledge, no one has been able to tie the knot 
on their equivalence and to derive the observed value unambiguously.

Finding an explanation of the value of $\rho_{\Lambda}$ has been
the focus of many research papers and projects. A somewhat favorite
approach nowadays is based on the weak anthropic principle, which
is consistent with the above demonstrated $\rho_{H}$ value. 
Thus, the Hubble constant $H_{0}$ is a parameter of the system while the
fundamental constants $c$, $G$, and $\hbar$ are really fundamental.
This way, the Planck length and time are ``unique'' once the Planck
system of units is chosen, while the Hubble constant $H_{0}$ is somehow
related to the details of how much stuff is in the system and what
are the initial conditions of the relevant processes. 
Systems near equilibrium often have a characteristic scale that is a result of
a few different competing factors. In physics, this is often reflected
by the kinetic and potential energy of a system or the balance of
various forces. In this respect, a reasonable approach 
that can link the Planck scale to the observed size of the Universe 
should then deliver a resolution to the Cosmological Constant Problem (CCP) 
above and may shed light on the problem of quantum gravity.

It is natural to expect that the value of the cosmological constant
would be determined or justified within Quantum Cosmology (QC) or
at least QC will provide some initial understanding of why it is so
small. The problem is that QC involves path integrals over 4D-geometries.
While path integrals seem to make sense in physics---they are not
yet well-defined mathematical objects. 
It seems, however, that progress in causal dynamical triangulations 
\cite{Mottola1995,Ambjorn2004} may provide a method to compute 
such complicated path integrals over 4D-geometries.

The focus of the current paper is the construction of a partition function
over a \textit{characteristic-size} variable that should be equivalent
to a path integral over 4D-geometries when all other characteristics
of the corresponding universe are integrated out. Such research is
motivated by the progress in causal dynamical triangulations 
\cite{Mottola1995,Ambjorn2004} and the possibility to numerically simulate various space-time geometries
for the purpose of testing quantum cosmology models. Thus, far in this
paper, the~Cosmological Constant Problem (CCP) has been introduced
and a few of the directions, which~researchers are exploring as resolutions
to this problem, have been briefly mentioned. The~next section outlines
five of the main categories as discussed by Weinberg \cite{Weinberg1989}
with a particular focus on the Weak Anthropic Principle and Quantum
Cosmology as well as an additional category of geometry related/motivated
approaches for justifying the value of $\rho_{\Lambda}$. Then, the
mathematical structure of Quantum Cosmology is briefly discussed along
with how the cosmological constant and the matter vacuum energy can
be absorbed in an effective cosmological constant. Once the Euclidean
partition function over geometries has been constructed, one can map
the path-integral partition function onto one-dimensional integral over the 
\textit{characteristic size} of each possible universe---\textit{Multiverse Partition Function}. Finally, the implications
of this partition function and possible interpretations of the 
\textit{characteristic size} of a universe are discussed.

\section{The Three Main Model Groups and the Seven Principal Model Categories}

The Planck scale is a well-accepted scale where something new and
interesting should start showing up; however, it may not be applicable
as a scale for estimating the vacuum energy--density fluctuations relevant
to the large-scale cosmological phenomenon. It may well be the same
reason for which a matter density of a material is correctly estimated
as macroscopic quantity and, if one uses sub-atomic scale measurements,
then the density of the material would exhibit significant variations.
Thus, without a suitable procedure/model for extrapolating from one
scale to another, the results may be drastically incorrect. As a result,
there are many different approaches based on various ideas as explanations
of the value and the origin of the non-zero cosmological constant. 

In a 1989 review paper, Weinberg discussed five main categories \cite{Weinberg1989}:
(a) Super-fields and super-potential approaches based on Supersymmetry,
Supergravity, and Superstrings; (b) Anthropic considerations associated
with processes in the Universe related to the observed age, mass density,
and~other astrophysical observations; (c) Adjustment/Tuning mechanisms
based on a scalar field and its evolution over the history of the
Universe; (d) Changing Einstein GR/ Modifying Gravity either by restricting the dynamical
degrees of freedom associated with the metric tensor and/or the relevant
symmetry transformations or by adding new terms to the Action that
result in a cosmological constant term as a constant of integration;
(e) Quantum Cosmology as pertained to estimating the probability of
observing a specific field configurations and relevant field expectation
values build on appropriate effective actions. 
{\it Each of these five categories can be grouped further} 
either as high-energy/short-distance scale phenomena (a) \& (c), 
or long-time/large-distance scale phenomena (b) \& (d), 
or presumably all scales (c, d, e). While the models from
the first two groups are self-evident and are expected to be extendable
into the all-scales group, these models are originally devised, developed,
and tested or expected to hold in their primary domain. The models
like (c, d, e) are built with the aim and hope to be applicable to
all scales.

The five Weinberg's categories are the natural classification based
on where the Occam's razor principle could lead us to in terms of
the minimal and sufficient change towards resolution of the CCP. The
minimal and sufficient change may come from extension to the Einstein's
GR theory of Gravity, category (d) above or to the theory behind the
matter fields that generate the stress--energy tensor, category (a)
above; the new theory, however, should be consistent with observations
about our Universe described in category (b). The successful model
will probably have an effective description in terms of the category
(c). Finally, if the model is to be a bridge towards fusing the principle
of general relativity and quantum mechanics that is applicable to
the Universe at a larger scale than it should in some sense be related
to quantum cosmology or may define what this category (e) means.

Models that have a mechanism of connecting short to long scale phenomena
are known in the literature, for example, T-duality in string theory,
or models that are the same at any scale, i.e., fractal or conformal
geometry \cite{Nottale2008,Lucat2018}. Since the physics deduced
from astronomical observations at large distances is expected to be
somehow related to the physics that can be studied at short distances,
for example in any on/near Earth laboratory, then such scale mapping
procedure should play a role in models that may resolve the CC problem.
In this respect, models that exhibit, scale invariance, conformal symmetry,
T-duality, or fractal geometry should be another (sixth) category
of models---(g) geometric models 
\cite{Pathria1972,Nottale2008,Lucat2018,Perelman2018a,Brans1961}.
Of course, there is always the possibility to consider models that
don't fit in any of these six categories the (seventh) category type---(f) further futuristic/exotic models category.

Relatively recently, the Weinberg's classification was discussed and
extended further in a review paper by Li {et al.} \cite{Li2011}
by adding more symmetry based models or string-theory motivated models
that either expand the matter sector, or modify gravity, or add further
examples to the geometry related model category (g) discussed in the
paragraph above. For example, Wetterich's scaling invariance via dilatation
field, or the Blackhole Self-Adjustment model that can be viewed as
a variation on the theme of Blackhole Cosmology models. The Holographic
principle category in Li's paper has a significant overlap with the
geometry related model category (g) which can also absorb the Back-Reaction
category in Li's paper. Some of the Phenomenological models (Quintessence,
K-essence, and so on) could also be viewed as members of the category
(c) adjustment/tuning mechanisms category by Weinberg, or~the (f)
category of futuristic models that expand the matter fields models
either in the particle physics sector or the type of fluid(s) (Chaplygin
gas, viscous fluid, super-fluid condensation) involved in the relevant
Freedman equations.

The fact that researchers have been exploring various models, like
string theory tuning mechanisms for bouncing brane-world scenarios,
Blackhole self-adjustment, Holographic principle, and Back-Reaction
models, is adding weight to the importance of the category (g) of
geometry related/motivated approaches for justifying the value of
$\rho_{\Lambda}$. If a model has a way to relate the short-scale
$r$ processes to large-scale $R$ processes, i.e., as the T-duality
in string theory, there will be a fundamental constant of dimension
$L^{2}$ such that ($r=L^{2}/R$), and therefore an energy density
associated with it ($\rho\sim c^{4}L^{-2}/(8\pi G)$). If this scale
is the string tension, as in string theory, then one still has an
enormous energy--density of the string vacuum compared to the observed
value. In a similar way, the Black-Hole related reasoning \cite{Li2011}
results in an energy--density expression ($\rho\sim3L^{-2}/(8\pi G)$)
that has the correct numerical factor of 3 which gives support to
the Blackhole Cosmology ideas where our Universe is considered to
be essentially the interior of a huge Blackhole \cite{Brans1961,Perelman2018a,Poplawski2018}.
Thus, this~idea could be used to explain the value of the observed
energy density, if one uses the future event horizon $L=R_{h}$, and
thus the remarkable agreement of dimensional argument (\ref{eq:rho_H0})
with the observation result (\ref{eq:rho_Lambda_exp}). 
If one considers the Blackhole Cosmology, then there is an alternative explanation of
the cosmological constant $\Lambda$ as related to the entropy of the Universe rather than
the energy--density of the vacuum. Such approach can use the entropy of a de Sitter Space
as derived by Gibbons and Hawking
%MDPI: Starting with the original Ref 24, the following documents were reordered, please confirm.
% VGG:Yes, it is ok!
~\cite{Gibbons&Hawking1977} $S_{dS}$
and match it to the corresponding expression for a blackhole  $S_BH$:
\begin{equation}
S_{dS}=12\pi\Lambda^{-1} \equiv S_{BH}=\frac{A}{4}
\label{eq:SdS}
\end{equation}
in units $\frac{k_B}{{l_P}^2}=1$, where $k_B$ is the Boltzmann constant, $l_P$ is the Planck length, 
and  $A=4\pi R^2$  is the area of the event horizon $R$ of the blackhole. 
The result is then $R=2l_{\Lambda}$, where $l_{\Lambda}=\sqrt{\frac{3}{\Lambda}}$
is the characteristic size of the de Sitter spacetime with a cosmological constant 
$\Lambda$ \cite{deSitter1917c}. This justifies the simple dimensional results 
discussed after (\ref{eq:rho_H0}) as related to the entropy of the Hubble horizon $R_H=\frac{c}{H_0}$
when employing results from path integral quantum gravity considerations
\cite{Gibbons&Hawking1977}. This relation between $H_0$  and $\Lambda$ 
is expected to be influenced minimally upon the specific details of the final quantum gravity theory 
because the relations discussed are about the entropy of  
the de Sitter spacetime which is the asymptotic limit of our Universe. 
Thus, while (\ref{eq:SdS}) could provide a new viewpoint on the smallness and the value of $\Lambda$,
the unimodular gravity approach to Einstein's equation \cite{Smolin2009}
could provide the resolution to the other two puzzles about the cosmological constant.
Alternatively, the smallness of  $\Lambda$ could be related to second-order quantum corrections
to the equations of motion either via the quantum corrected Raychaudhuri equation \cite{Farag&Ali&Das2015}, 
or as a second-order quantum correction to the Newtonian gravity via the use of 
the quantum hydrodynamic approach \cite{Chiarelli2019, Chiarelli2020}.
Such an approach is very appealing due to its economy in terms of exotic new stuff and as a member 
of the category (d) above where the cosmological constant is related to the quantum vacuum fluctuations
that induce effectively conformal effects via the non-zero trace of the corresponding energy--momentum 
matter tensor.

The numerical agreement between (\ref{eq:rho_Lambda_exp}),  (\ref{eq:rho_H0}), and (\ref{eq:SdS}),
however, is not sufficient to establish the validity of a model
or the idea behind it as outlined in the example above. In order to do so, one has to build a model
that can provide a description to the vast body of observational data
to the level compatible with what the $\Lambda$CDM model can provide.
Recently, a model based on the idea of scale-invariance of the vacuum (SIV)
and the Weyl's conformal gravity idea has been in development and
has been showing a promise to resolve the dark-matter problem dynamically
and possibly without the need for dark-matter particles \cite{Maeder2017b,Maeder2017}. 
Connecting the SIV theory to the unimodular gravity may be a fruitful approach
to resolving the various dark puzzles in cosmology and gravity. 
Further fortification of the argument for such connection can be justified 
given the fact that the unimodular gravity can be viewed as a particular gauge fixing  
of the conformal factor as a constraint \cite{Mazur&Mottola1990}. In this respect, the
specific SIV realization of the Weyl Integrable Geometry with suitable conformal factor 
is equivalent to the Einstein GR equation with traceless energy--momentum for matter  and 
a Cosmological Constant term originating from a non-flat background characterized 
by non-zero Ricci scalar which absorbs the over all trace contribution of 
the matter energy--momentum tensor. Such non-flat Lorentzian backgrounds are expected to be
viable stable vacuum configurations for quantum gravity \cite{Mazur&Mottola1990}.
The~studies by Mottola and his collaborators seem to be intimately related to the 
understanding of the trace of the stress--energy tensor (\ref{eq:Tand_Lambda}), 
and thus to the value of $\Lambda$ as induced by the ``energy'' of the gravitational field.
The study of the trace-anomaly-induced dynamics of the conformal factor of 
four-dimensional (4D) quantum gravity \cite{Antoniadis&Mottola1992} seems to be
closely related to the scale-invariance of the vacuum (SIV) and the Weyl's conformal gravity idea
explored by the authors of the current paper in a different context~\cite{MaedGueor19}. 
Unfortunately, despite the simple relation between the field contents of these studies 
($\sigma=\ln{\lambda}$), we couldn't finalize the correspondence on   
the level of the equation content within the short time frame 
we have been studying Mottola's work for the current paper.
Establishing the connection is an important step and has to be done
carefully and accurately, which usually takes significant time, effort, and
other relevant resources.

Models that successfully compete with the $\Lambda$CDM model are
effectively satisfying the Weak Anthropic Principle. That is, the
parameters of the model do allow for good description of the observed
physical reality, even though part of the parameter space may not
be consistent with the existence and evolution of Life as we know
it. Current understanding of the inflationary Universe model suggests
that there could be the possibility of a Multiverse where various
universes are perpetually created as the inflation field expands and decays locally. 
As a result, the laws of physics may differ in different universes and 
we just happen to live in a Universe that has the observed structure. 
When the anthropic reasoning is properly utilized, 
it can help us understand the correlations between various observational facts and 
the corresponding numerical representations of the data 
\cite{Weinberg1989,Vilenkin2001}.

An alternative multiverse scenario is the Blackhole Cosmology
idea where each Universe is a Blackhole that is either a daughter
Universe or parent Universe. This way the Blackhole universes are
disconnected universes, but the laws of physics are close in the generationally
related universes. The~above multiverse options are interesting due
to their implication about the overall history and fate of the Universe
and therefore are more consistent with viewing the cosmological constant
as constant of integration related to the initial conditions of the
system or a byproduct of a specific cancellation~mechanism. 

An alternative equally radical view point is based on the idea of self-similarity
and scale relativity as related to fractal geometry \cite{Nottale2008} 
where the value of the cosmological constant is related to the scale of
the classical electron radius, thus a byproduct of a specific phase
transition mechanism. Regardless of the mechanism and the origin of
the specific value of the cosmological constant, there is always a scale
that should characterize a quasi-static cosmological constant value.

Quantum Cosmology (QC) is an approach aiming at understanding the
Universe at the larger scale by using the ideas behind general relativity
and quantum mechanics. It has been used previously to justify that
the cosmological constant is probably zero. Some initial arguments
were based on supersymmetry \cite{Hawking1983} and quantum gravity 
\cite{Coleman1988}, while reparametrization invariance considerations
were used to demonstrate that the cosmological constant can be viewed
as an integral of the motion \cite{Weinberg1989}. 
The~QC approach was viewed as the most promising road back then, 
but the experimental observation has raised a red flag over this method.
Some QC studies argued that the zero Cosmological Constant result is not 
that much a statement about what should be observed as the most likely value
of $\Lambda$ but more of a statement about the asymptotic ``far future'' value
of the model parameter $\Lambda$ corresponding to model based on Baby Universes 
\cite{Kawai&Okada2011}.
Another argument in support for the zero Cosmological Constant 
as related to the asymptotic infrared stable fixed point for 4D gravity has been given
very rigorously and much earlier then in \cite{Kawai&Okada2011}
by Mottola and his collaborators \cite{Antoniadis&Mottola1992,Mottola1995}.

Subsequent adjustments to the Multiverse approach seem to be aligned with
observational support for inflation. In the next few sections, an approach
based on the isolation of a characteristic scale as an integration
parameter is considered for the purpose of estimating the 
expected average size of ``typical'' universe as part of a Multiverse ensemble.

\section{Cosmological Constant within Quantum Cosmology}

The Einstein equations with cosmological constant $\Lambda$: 
\begin{equation}
R_{\alpha\beta}-\frac{1}{2}g_{\alpha\beta}(R-2\Lambda)=\frac{8\pi G}{c^{4}}T_{\alpha\beta}\label{eq:EFE}
\end{equation}
can be derived from the Einstein--Hilbert action supplemented with the matter action $A_{matter}$: 
\begin{equation}
A[g,\psi]=\frac{c^{4}}{16\pi G}\int(R-2\Lambda)\sqrt{-g}d^{4}x
+A_{matter}[g,\psi,\partial\psi].\label{eq:TheAction}
\end{equation}

The first term above is the standard Einstein--Hilbert action for a metric gravity $g_{\alpha\beta}$ 
with cosmological constant $\Lambda$ while the second term is the matter action $A_{matter}$ 
that depends on the metric $g_{\alpha\beta}$ but not on its derivatives and thus it determines 
the stress--energy tensor $T_{\alpha\beta}$ but does not contribute to the left-hand side of (\ref{eq:EFE}). 
In $A_{matter}$, the matter fields are denoted by $\psi$ and their derivatives are $\partial_{\mu}\psi$. 
The symbol $\psi$ is a placeholder for the variety of possible matter fields like those in 
the standard model of elementary particles, such as leptons, quarks, and interaction bosons.
The equations obeyed by the matter fields within given background spacetime with a metric $g_{\alpha\beta}$ 
are obtained upon the variation of the relevant fields when the 
appropriate functional form of $A_{matter}$ is considered.

The expectation value of an observable $\mathcal{O}[g,\partial g,\psi,\partial\psi]$
is then defined via a partition function that is a formal Feynman
path integral with a functional integration measure $\mathcal{D}[g,\psi]$ 
over the space of varius possible and relevant metrics and matter fields: 
\begin{eqnarray*}
\left<\mathcal{O}\right> & = & \frac{1}{Z}\int\mathcal{D}[g,\psi]\mathcal{O}[g,\partial g,\psi,\partial\psi]e^{\frac{i}{\hbar}A[g,\psi]},\\
Z & = & \int\mathcal{D}[g,\psi]e^{\frac{i}{\hbar}A[g,\psi]}.
\end{eqnarray*}

When the matter fields $\psi$ are integrated out, one expects $T_{\alpha\beta}$
to be proportional to the metric tensor $g_{\alpha\beta}$ which leads
to an effective $\Lambda$ for an empty Universe \cite{Weinberg1989, Pejhan2018}:
\begin{equation}
\left\langle T_{\alpha\beta}\right\rangle =
\sigma g_{\alpha\beta},~\Lambda\rightarrow\Lambda-\frac{8\pi G}{c^{4}}\sigma.
\label{eq:Tand_g}
\end{equation}

Here, $\sigma$ denotes the constant of proportionality between 
$\left\langle T_{\alpha\beta}\right\rangle $ and the metric tensor $g_{\alpha\beta}$. 
Therefore, it is related to the effective energy--density $\left\langle T_{00}\right\rangle $
when the quantum fields $\psi$ are integrated out. If one is to consider
the trace of the effective energy--momentum tensor $\left\langle T_{\alpha\beta}\right\rangle $,
then one has $4\sigma=\left\langle T_{\alpha\beta}\right\rangle g^{\beta\alpha}$.
It seems reasonable to assume that the trace operation commutes with
the process of integrating out the fields $\psi$ which leads to the relationship 
$4\sigma=\left\langle T_{\alpha\beta}g^{\beta\alpha}\right\rangle =\left\langle T\right\rangle $,
where $T=T_{\alpha\beta}g^{\beta\alpha}$ is the trace of the energy--momentum tensor. 
If one chooses traceless formulation for the Einstein GR equations, then the $\Lambda$
appears as constant of integration related to the Ricci scalar curvature $R$ and the trace $T$
of the energy--momentum tensor \cite{Weinberg1989}.

In units where the speed of light is $c=1$ and
the metric tensor has time-like signature $\left\{ +,-,-,-\right\}, $
the energy--momentum tensor for an ideal fluid is given by the
expression: $T^{\alpha\beta}=(\rho+p)u^{\alpha}u^{\beta}-pg^{\alpha\beta}$.
Here, $\rho$ is the energy--density, $p$ is the pressure, and
$u^{\alpha}$ are the components of the for-velocity of the fluid
with $u^{\alpha}u^{\beta}g_{\alpha\beta}=1$ and $u^{0}\approx 1$
in the Newtonian limit of a slow motion. Then, the trace of the energy--momentum
tensor becomes $T=g_{\alpha\beta}T^{\beta\alpha}=(\rho-3p)$, which
turns into $T=\rho(1-3w)$ if an equation of state $p=w\rho$
is considered.

Based on the conservation laws in the scale-invariant theory, which are also
reduced to the usual conservation laws \cite{Maeder2017b}, the overall 
energy--density scaling $\rho\sim r^{-3(w+1)}$ indicates five special values of $w$.
The five values can be grouped as follows, three of which are particularly
important: radiation $(w=1/3)$, pressureless matter $(w=0),$ and
dark energy $(w=-1).$ The first two, radiation and matter, are familiar
non-negative pressure fluids. The negative $w$ results in negative
pressure systems that are not common laboratory fluids. The other
two special cases, $w=-1/3$ (string defects) and $w=-2/3$ (domain wall defects) 
\cite{Sahni2000}, correspond to $r^{-2}$ and $r^{-1}$ scaling, 
while the dark energy $(w=-1)$ corresponds
to constant energy density independent of $r$; 
thus, it does not scale with the size $r$ of the system for $w=-1$.
This is also the equation of state of the locally Lorentz invariant vacuum.

It is interesting to notice that the dark-energy equation of state
$(p=-\sigma)$ is implied by the assumption $\left\langle T_{\alpha\beta}\right\rangle =\sigma g_{\alpha\beta}$,
which cannot be satisfied neither by radiation nor by pressureless
dust alone. The trace of the radiation energy--momentum tensor is zero
$(T=\rho(1-3w)=0)$ due to $w=1/3$ for radiation; therefore, radiation
cannot contribute to the value of $\sigma$. 
If the energy--momentum tensor is traceless, then $\sigma=0$ and it would have 
implied $w=1/3$ (relativistic matter only) and $\Lambda$ would be related to 
the Ricci scalar $R$. Effectively, this is the case of the very,very early Universe 
where matter and radiation were in the state of very-hot quark--gluon plasma. 
However, the current Universe is far from such state.

If one assumes that locally $w\geq0$, then integrating out the matter fields $\psi$ will result in 
$4\sigma=\left\langle T\right\rangle 
=\left\langle \rho\right\rangle -3\left\langle w\rho\right\rangle <\left\langle \rho\right\rangle $,
which implies $\sigma/\left\langle \rho\right\rangle <1/4$. Thus,
the effective dark energy--density $\sigma$, of~radiation and matter, cannot
be more than a quarter of the overall total energy--density 
$\left\langle T\right\rangle \sim\left\langle \rho\right\rangle $
when $\left\langle w\rho\right\rangle \ll\left\langle \rho\right\rangle $.
Notice that this provides an interpretation of the dark-energy as
an effect arising from the integration of the matter fields and possible
cancellation mechanism for bringing the effective energy--density 
$4\sigma=\left\langle \rho\right\rangle -3\left\langle w\rho\right\rangle $
arbitrarily close to zero depending on the overall distribution of
$\rho$ and the relevant local values of $w\geq0$. 
Unfortunately, the inequality $\sigma<\left\langle \rho\right\rangle /4$ implies
that the energy--density of the dark-energy field $\sigma$ is less
then the radiation and matter energy--density $\left\langle \rho\right\rangle $ when $w\geq0$,
which contradicts the current state of the observed Universe. 
Of course, for the dark-energy field when $w=-1$, one
has $4\sigma=\left\langle T\right\rangle =4\left\langle \rho\right\rangle $,
thus, $\sigma=\left\langle \rho\right\rangle $ while radiation
and matter can be added on top of it as perturbations and one can consider 
$R=4\Lambda_{0}$. 
However, if one insists on $R=4\Lambda_{0}=0$ but without dark energy ($w=-1$), 
then $-1<w<0$ is needed to have $\sigma > \left\langle \rho_{\rm rad+matter}\right\rangle$, 
but it is still not enough to agree with observations. 
For example, given that $\Omega_\Lambda/\Omega_M=0.72/0.28\approx 2.6$,
to~have $\sigma$ to be about two to three times 
as much as $\left\langle\rho_{\rm matter}\right\rangle$ with
three species $w \in \{0, -1/3, -2/3\}$ 
each with an equal amount of energy--density $\rho$, 
one can only get to $\sigma/\rho=6/4=1.5$
since $4\sigma=\left\langle T\right\rangle 
=\rho+\rho(1-3\times(-1/3))+\rho(1-3\times(-2/3))=\rho\times(1+2+3)$.
\footnote{The ratio is even smaller ($\sigma/\rho=5/4=1.25$) 
if one is to consider equal energy partition per degree of freedom ($1/2$) 
with dimension degree of freedom $n$ deduced from the scaling $\rho\sim r^{-n}$.}
Note that dark-matter has been already included by considering $\Omega_M=0.28$.
Given the lack of laboratory observations of dark species obeying $-1<w<0$, 
a non-zero cosmological constant $\Lambda_{0}$ must be present to explain
the cosmological observation. Therefore, there must be non-zero Ricci curvature 
$R=4\Lambda_{0}$ associated with it.
Such consideration is consistent with viewing the effective cosmological
constant as a constant of integration \cite{Weinberg1989}:

\begin{equation}
\Lambda\rightarrow\Lambda_{eff}=\Lambda_{0}-\frac{8\pi G}{c^{4}}\sigma=
\frac{1}{4}R-\frac{8\pi G}{c^{4}}\frac{1}{4}\left\langle 
T_{\alpha}^{\alpha}\right\rangle .\label{eq:Lambda_eff}
\end{equation}

Therefore, since empty space (the vacuum) is dominating the Universe,
then one can integrate the matter fields and use an effective $\Lambda$
that has an implicit contribution from the matter (\ref{eq:Lambda_eff}). 
For flat space-time ($R=0$), this would imply that the value of the effective cosmological constant 
should be due to the trace part of the stress energy--momentum tensor $T_{\alpha\beta}$ and
should somehow be deducible from the  matter distribution. 
Then, the smallness of $\Lambda_{eff}$ would imply that it is a perturbative effect rather than
a leading order effect as implied by (\ref{eq:rho_Planck}).
In fact, models based on estimates of the quantum fluctuations using quantum potentials and 
quantum hydrodynamic formalism are pointing to such behavior when one is looking at the 
quantum corrected Raychaudhuri equation \cite{Farag&Ali&Das2015}, as well as 
the Classical Klein--Gordon Field \cite{Chiarelli2019}, and 
the Spinor-Tensor Gravity of the Classical Dirac Field \cite{Chiarelli2020}.
If~one adopts the above viewpoint  as the main sources of the non-zero vacuum energy density,
then one can accept the zero cosmological constant argument based on ideas from the 1980s and 1990s,
and then add a perturbative effect of sparse matter distribution on top of it.

However, from the point of view of the usual QC, the effective cosmological constant $\Lambda_{eff}$, 
the~origination of the non-zero value is irrelevant since flat spacetime ($R=0$) as well as
non-flat spacetime ($R\ne0$) are viable stable Lorentzian background vacua 
for  path-integral based Quantum Gravity models \cite{Mazur&Mottola1990}.
Thus, in (\ref{eq:EFE}) and (\ref{eq:TheAction}), one can set $T_{\alpha\beta}=0$
and $A_{matter}=0$, which effectively is the vacuum Einstein field
equation (EFE) with the corresponding action: 
\begin{eqnarray}
 & & R_{\alpha\beta}-\frac{1}{2}g_{\alpha\beta}(R-2\Lambda)=0\label{Einstein Equations}\\
 & & A[g]=\frac{c^{4}}{16\pi G}\int(R-2\Lambda)\sqrt{-g}d^{4}x\label{EH Action for GR}
\end{eqnarray}

Most modern vacuum cosmology models are derived from (\ref{Einstein Equations})
for specifically chosen metric tensor $g$. For homogeneous and isotropic Universe,
the Ricci scalar curvature is expected to be constant  or at most  time dependent only.
If one considers a vacuum solutions to (\ref{eq:EFE}), then one arrives at (\ref{Einstein Equations}),
which implies that $R=4\Lambda$. Quantum cosmology considers 
all reasonable possibilities for the metric tensor $g$ within the appropriate path
integrals  \cite{Gibbons&Hawking1977, Smolin2009,Antoniadis&Mottola1992,Mottola1995}.

It is usually assumed that (\ref{Einstein Equations}) provides a
satisfactory description of the physical reality at cosmological scales.
Therefore, the main contribution to the action (\ref{EH Action for GR})
comes from the ``classical trajectories'' satisfying (\ref{Einstein Equations}).
In this respect, one can approximate the action (\ref{EH Action for GR})
by considering an integration of (\ref{Einstein Equations}) and its
implication for (\ref{EH Action for GR}): 
\begin{eqnarray}
\int\left(4\Lambda-R\right)\sqrt{-g}d^{4}x & = & 0\label{R_eq_4Lambda}\\
A[g]\approx \frac{c^{4}}{16\pi G}\int2\Lambda\sqrt{-g}d^{4}x & = & \int\rho_{\Lambda}\sqrt{-g}d^{4}x
\label{A-classical}
\end{eqnarray}

One can justify the partition function $Z$ in the familiar form (
$e^{-\beta E}$ ) by switching to imaginary time, Euclidean metric
$g_{E}$, and, assuming the separation of the four-volume integral
into time integral and three-volume integral by using the following
relationships: 
\begin{eqnarray}
e^{\frac{i}{\hbar}A[g]} & \rightarrow & e^{-\beta E},~\beta=-\frac{i}{\hbar}\int dt,\nonumber \\
\sqrt{-g}d^{4}x & \rightarrow & dtdV^{(3)},~E=\int\rho_{\Lambda}dV^{(3)}.
\label{eq:switching_to_imaginary_time}
\end{eqnarray}

In view of the above expressions, one expects that the partition function
for a set of metrics $g$ satisfying (\ref{Einstein Equations}) is
given in terms of the corresponding Euclidean metric $g_{E}$ \cite{Weinberg1989}:
\begin{equation}
Z=\int\mathcal{D}[g_{E}]\exp\left(-\frac{c^{4}}{8\pi G\hbar}\int\Lambda\sqrt{g_{E}}d^{4}x\right).
\label{PartitionFunction}
\end{equation}

When evaluated using a Hartle--Hawking no-boundary condition in complexified  spacetime, 
one gets to 4-spheres of radius $R=\sqrt{3/\Lambda}$ and obtains a probability 
density distribution \cite{Weinberg1989, Gibbons&Hawking1977,Hawking1984b}:
\begin{equation}
P\sim e^{-I_{eff}}=\exp{(\frac{3\pi c^{4}}{G\hbar\Lambda})}
\label{zeroCC}
\end{equation}

Based on this expression of the probability density distribution, Hawking has
concluded that the cosmological constant is probably zero \cite{Weinberg1989,Hawking1984b}.
Various approaches have been  suggested on improving the probability 
density distribution since the measured value of $\Lambda$ has turned out to be very small but non-zero.
One of the most current approaches have been focused on the 4-volume cutoff measure of the multiverse \cite{Vilenkin2020}.
The discussion in \cite{Hawking1984b} considers also an anti-symetric tensor $A_{\mu\nu\gamma}$ 
that is one way to arrive at unimodular gravity \cite{Weinberg1989,Smolin2009}.
The above result (\ref{zeroCC}) relies on the observation that 
$\int\sqrt{g_{E}}d^{4}x=\nu\left[g_{E}\right]\zeta^{4}$
where the characteristic size $\zeta$ is actually the 4-spher radius $R=\sqrt{3/\Lambda}$.
The~overall geometric dimensionless factor of proportionality is denoted by 
$\nu\left[g_{E}\right]$. For example, this~factor is $1$ for standard Euclidian n-cube, 
and $4\pi/3$ for the volume of a 3D sphere. Calculations for the tunneling probability 
from ``nothing'' into something with a 4D size $R=\sqrt{3/\Lambda}$ 
considering wormholes, using Euclidian action and imaginary time, 
have resulted in the expression $P\sim\exp(1/\Lambda)$,
while the use of real time and WKB-approximation result in $P\sim\exp(-1/\Lambda)$ instead 
\cite{Coleman1988, Kawai&Okada2011}. It has been argued that the more appropriate probability 
expression is  $P\sim\exp(-1/{\left|\Lambda\right|})$ \cite{Vilenkin1984}.
The sign difference can be traced to whether there is a minimum size or maximum size 
of the 4D manifold as well as how the overall path integral measure is handled
\cite{Mazur&Mottola1990}.
In the forthcoming discussion, we will utilize (\ref{PartitionFunction}) 
and will not impose such relationship ($\zeta \sim R=\sqrt{3/\Lambda}$) 
between characteristic size $\zeta$ and $\Lambda$.
Our~goal is to estimate the typical size of an universe within the multiverse approach.
The~calculations discussed in the next section show that the typical size of an universe is of the order 
of the Planck scale and therefore the Quantum Field Theory estimates of the vacuum energy density 
(\ref{eq:rho_Planck}) is a reasonable~expectation.

\section{Multiverse Partition Function and the Typical Size of a General $\zeta$-Universe}

Consider an ensemble of universes (Multiverse) where each universe
is a solution to the Einstein field Equation (\ref{Einstein Equations})
and is labeled by $\{g_{\alpha\beta},\Lambda,\zeta\}$. Here, $\zeta$
is an auxiliary tag, a characteristic size, to~a given EFE solution
$\{g_{\alpha\beta},\Lambda\}$. The characteristic size can be determined
by measuring the biggest distance between two spacial points, or the
age of the Universe $t=\zeta/c$, or its 4D volume 
$\int\sqrt{g_{E}}d^{4}x=\nu\left[g_{E}\right]\zeta^{4}$,
at a particularly important moment of its evolution and so on. 
The definition of $\zeta$ is loose and fuzzy in order to
fulfill its purpose to be an enumerator for various  
solutions of the Einstein field equations.
In order to be able to set up the integration variables in the partition
function in a manageable form, one can introduce an equivalent set
of fields and variables: 
\begin{eqnarray}
\{g_{\alpha\beta},\Lambda,\zeta\}\rightarrow\{\tilde{g}_{\alpha\beta},\tilde{\Lambda},\tilde{\zeta} :a\}
\end{eqnarray}
where $g_{\alpha\beta}=a^{2}\tilde{g}_{\alpha\beta}$
the second power of $a$ is a matter of convenience; 
$a$ is simply related to $\zeta$ since the distance 
between two points along an $s$-parametrized curve is: 
\begin{eqnarray}
\zeta=\int ds\sqrt{g_{\alpha\beta}\frac{dx^{\alpha}}{ds}\frac{dx^{\beta}}{ds}}
=a\int ds\sqrt{\tilde{g}_{\alpha\beta}\frac{dx^{\alpha}}{ds}\frac{dx^{\beta}}{ds}}
\end{eqnarray}

Thus, $\zeta=a\tilde{\zeta}$. The choice $\tilde{\zeta}=1$ is particularly
important and will be discussed later. 

Furthermore, notice that, if $\{g_{\alpha\beta},\,\Lambda\}$ solves the Einstein field Equation
(\ref{Einstein Equations}), then $\{\tilde{g}_{\alpha\beta},\,\tilde{\Lambda}\}$
would also be a solution to the Einstein field Equation (\ref{Einstein Equations})
where $\Lambda$ and $\tilde{\Lambda}$ are simply related as $\Lambda=\tilde{\Lambda}/a^{2}$.
This follows from the fact the affine connection $\Gamma_{\beta\gamma}^{\alpha}$
is invariant under rescaling of the metric tensor and so are the Riemann
curvature tensor $R_{\beta\gamma\delta}^{\alpha}$ and the Ricci curvature
tensor $R_{\beta\delta}=R_{\beta\alpha\delta}^{\alpha}$, but the scalar curvature 
$R=g^{\alpha\beta}R_{\alpha\beta}$ rescales as $g^{-1}$ which is effectively $a^{-2}$.
One simple way to see that the affine connection $\Gamma_{\beta\gamma}^{\alpha}$
is invariant under rescaling of the metric tensor $(g_{\alpha\beta}=a^{2}\tilde{g}_{\alpha\beta})$ 
is to notice that $\Gamma_{\beta\gamma\alpha}$ is built by using partial derivatives of the metric
tensor $g_{\alpha\beta}$; thus, $\Gamma_{\beta\gamma}^{\alpha}=g^{\alpha\nu}\Gamma_{\beta\gamma\nu}$
where $g^{\alpha\nu}$ is the inverse of $g_{\alpha\nu}$ and therefore it scales with $a^{-2}$; 
that is, $g^{\alpha\beta}=a^{-2}\tilde{g}^{\alpha\beta}$.
Another way to justify $\Lambda=\tilde{\Lambda}/a^{2}$ is to recall that $T_{\alpha\beta}$ 
is scale invariant \cite{Maeder2017b}, which can be guaranteed only if $\Lambda=\tilde{\Lambda}/a^{2}$ when 
$g_{\alpha\beta}=a^{2}\tilde{g}_{\alpha\beta}$ given the Equations 
(\ref{eq:Tand_g}) and (\ref{eq:Tand_Lambda}).

Given an ensemble of solutions $\{g_{\alpha\beta},\Lambda,\zeta\}$,
where $\zeta$ has been determined by using a particular measurement
procedure/definition, one can split the ensemble into same-scale $\zeta$-subset,  for example 
$\{g_{\alpha\beta},\Lambda,\zeta=1\}\equiv\{\tilde{g}_{\alpha\beta},\tilde{\Lambda},\tilde{\zeta}=1:a=1\}$
members, and $a$-derivable members, such that 
$\{g_{\alpha\beta},\Lambda,\zeta\}\equiv\{a^{2}\tilde{g}_{\alpha\beta},\tilde{\Lambda}/a^{2},a\tilde{\zeta}:\tilde{\zeta}=1\}$.
Let's call the $\zeta=\tilde{\zeta}=1=a$ subset---{\it the Planck-scale seed universes subset.}

The goal now is to write a partition function $Z$ in terms of the
variables $(\tilde{g}_{\alpha\beta},\,\tilde{\Lambda},a)$ by using
guidance from $Z[g_{E}]$ (\ref{PartitionFunction}). 
By following the mapping outlined in (\ref{eq:switching_to_imaginary_time}), one has: 
\begin{eqnarray}
\mathcal{D}[g_{E}] & & \mathrel{\mathop{\longrightarrow}
\limits _{g_{\alpha\beta}\rightarrow a^{2}\tilde{g}_{\alpha\beta}}}d[a^{2}]\mathcal{D}[\tilde{g}_{E}]\\
\int\Lambda\sqrt{g_{E}}d^{4}x & & \mathrel{\mathop{\longrightarrow}\limits _{g_{\alpha\beta}
\rightarrow a^{2}\tilde{g}_{\alpha\beta}}}\int\frac{\tilde{\Lambda}}{a^{2}}\times a^{4}\sqrt{\tilde{g}_{E}}d^{4}x.
\end{eqnarray}

Since each $a$-derivable member is built from a particular $\tilde{\zeta}=1$
solution $\{\tilde{g}_{\alpha\beta},\tilde{\Lambda},\tilde{\zeta}=1\}$,
then its contribution to the functional integrals is easily taken
into account through the variable $a$. 
Therefore, it~seems reasonable to define $Z$ and $\left\langle \mathcal{O}\right\rangle $
in Planck units ( $c=G=\hbar=1$ ) in the following way: 
\begin{eqnarray}
\label{Z(Lambda)}
Z & = & \int_{0}^{\infty}d[a^{2}]\int\mathcal{D}[\tilde{g}] 
e^{-\frac{1}{8\pi}a^{2}\tilde{\Lambda}\tilde{\nu}\tilde{\zeta}^{4}},\label{Z_zeta_a2} \\
\label{ExpValue}
\left\langle \mathcal{O}\right\rangle & = & \frac{1}{Z}\int_{0}^{\infty}d[a^{2}]
\int\mathcal{D}[\tilde{g}]\mathcal{O}e^{-\frac{1}{8\pi}a^{2}\tilde{\Lambda}\tilde{\nu}\tilde{\zeta}^{4}}.\label{<zeta>}
\end{eqnarray}

Here, $\nu[g_{E}]$ is an overall geometric factor that encodes most
of the geometric information related to $g.$ Usually, $\nu[g_{E}]$ is of order one. 
For example, for $R^{4}$ Euclidean geometry $\nu=1$,
for $D^{4}$ geometry $\nu=\frac{\pi^{2}}{2}\approx 5.$ However, for
a spherical blackhole with $8\pi M$ period in the imaginary time
and event horizon at $R=2M$, one has $\nu=\frac{16}{3}\pi^{2}\approx 53$. 
In general, one can consider classes of metrics $g_{E}$ for which 
$\nu[g_{E}]$ has the same value. Then, the ``path-integral'' over that class
$\mathcal{D}[\tilde{g}]$ would result in the ``volume'' $\mathcal{V}[\nu[g_{E}]]$
of this class of metrics and $\nu[g_{E}]$ can be used as a new integration variable. 
Thus, the Jacobian for this change of variables would be the metric volume
$\mathcal{V}[\nu[g_{E}]]$.
Since, as already mentioned, $\nu[g_{E}]$ is of order one usually; 
thus, one can neglect, for the present study, its dependence on the metric. 
However, in the future, one may consider $\nu[g_{E}]$ as an independent 
degree of freedom along which one would have to integrate.

Since the Planck-scale subset consists of Planck-scale universes,
it is plausible that at such sub-microscopic scale the quantum field
theory estimates of the vacuum energy density is appropriate; that
is, $\rho_{\tilde{\Lambda}}$ is of order 1. Thus, if for each solution
$\{g_{\alpha\beta},\Lambda,\zeta\}$ there is Planck-scale seed Universe
such that $a$ is chosen so that $\tilde{\zeta}=1$, $\tilde{\Lambda}$
in the range $(1,8\pi)$, which is compensating for fixing $\tilde{\nu}\approx 1$, 
and~$\mathcal{O}=\zeta=a\tilde{\zeta}=a$; then, evaluating (\ref{Z_zeta_a2}) and (\ref{<zeta>}) 
results in the average characteristic size $\left\langle{\zeta}\right\rangle $
of this $a$-derivable subset to be of order one as well 
($\left\langle {\zeta}\right\rangle \in(2\pi,\sqrt{\pi/2})\approx (6.3,1.25)$
when $\tilde{\Lambda}\in(1,8\pi)$).
Note~that, according to (\ref{Z(Lambda)}) and (\ref{ExpValue}), one has 
$Z\sim{\tilde{\Lambda}}^{-1}$, while 
$\left\langle\zeta \right\rangle\sim{\tilde{\Lambda}}^{-1/2}$.

One can easily see this upon changing the order of the integration and by performing 
the following substitution that makes the integration over $[a^2]$ very easy numerically
by considering the Plank-seed universe class $(\tilde{\zeta}\approx 1, \tilde{\nu}\approx 1)$
that results in a universe of characteristic size $\zeta=\tilde{\zeta}a$. 
Thus, the argument of the exponential function can be written from 
$\frac{1}{8\pi}(a^{2}\tilde{\zeta}^{2})(\tilde{\Lambda}\tilde{\nu})(\tilde{\zeta}^{2})$
into $\frac{1}{8\pi}(\zeta^{2})(\tilde{\Lambda}\tilde{\nu})(\tilde{\zeta}^{2})=\zeta^{2}\tilde{c}^2$.
We~will be performing the integration over $\zeta^{2}$ while keeping $\tilde{\zeta}=1$
and assuming $\tilde{\nu}\approx 1$ and $\tilde{\Lambda}$ in the range $(1,8\pi)$,
which is consistent with the Plank-seed universe class idea, while 
$\tilde{c}^2=\frac{1}{8\pi}(\tilde{\Lambda}\tilde{\nu})(\tilde{\zeta}^{2})$.
\begin{eqnarray}
\label{ZasLambda}
Z & = & \int\mathcal{D}[\tilde{g}] \int_{0}^{\infty}d[a^{2}] 
e^{-\frac{1}{8\pi}(a^{2}\tilde{\zeta}^{2})(\tilde{\Lambda}\tilde{\nu})(\tilde{\zeta}^{2})}
= \int\mathcal{D}[\tilde{g}] \int_{0}^{\infty}d[\zeta^{2}] \tilde{\zeta}^{-2} e^{-\zeta^2\tilde{c}^2}
= \frac{8\pi}{\tilde{\Lambda}} \int\mathcal{D}[\tilde{g}]\nu[\tilde{g}]^{-1},
\\
\label{zeta1}
\left\langle \zeta\right\rangle & = & 
\frac{1}{Z}\int\mathcal{D}[\tilde{g}] \tilde{c}^{-3} \int_{0}^{\infty}d[\zeta^{2}] {\zeta} e^{-\zeta^2}
=\frac{1}{Z}\int\mathcal{D}[\tilde{g}] \tilde{c}^{-3} \sqrt{\frac{\pi}{2}} 
=\frac{16\pi^2}{Z \tilde{\Lambda}^{3/2}}\int\mathcal{D}[\tilde{g}] \nu[\tilde{g}]^{-3/2}\approx  
\frac{2\pi}{\sqrt{\tilde{\Lambda}}}.
\end{eqnarray}
where we have used $\nu[\tilde{g}]=1$ assuming we can move the actual value 
of  $\nu[\tilde{g}]\approx 1$ into the value of  $\tilde{\Lambda}$ without that changing 
the range $\tilde{\Lambda}\in(1,8\pi)$ significantly but only rearranging the elements there.

In this respect, any particular solution $\{g_{\alpha\beta},\Lambda,\zeta\}$
of the EFE could be viewed either as $a$-derivable element connected
to a particular Planck-scale seed solution or it could be a Planck-scale
seed solution itself. This means that the multiverse ensemble can
be enumerated by Planck-scale seed universes and their $a$-derivable
elements.

The scale of the observed Universe is clearly far larger than the
Planck-scale, thus our Universe is $a$-derivable Universe based
on some Planck-scale seed Universe 
$\{\tilde{g}_{\alpha\beta},\tilde{\Lambda}\in(1,8\pi),\tilde{\zeta}=1\}$
so one can expect its characteristic size to be related to its present
size, which is $r=10^{26}~\text{m}$ in Planck units $(l_{P}=1.6\times10^{-35}~\text{m})$; 
this gives $a\approx 6\times10^{60}$ and therefore 
$\Lambda=\tilde{\Lambda}/a^{2}=8\pi/(6\times10^{60})^{2}=7\times10^{-121}$
or $\Lambda=\tilde{\Lambda}/a^{2}=1/(6\times10^{60})^{2}=3\times10^{-122}$
in Planck units. These values are clearly much closer to the measured
value $\rho_{\Lambda}/\rho_{Planck}=1.4\times10^{-123}$.

\section{Discussions and Conclusions}

One may expect further improvements in $\Lambda$ by adding quantum
effects via higher order terms in the Einstein--Hilbert action that
are beyond the approximation (\ref{A-classical}),
which considers only classical fields satisfying (\ref{Einstein Equations}),
and by making more rigorous calculations using $Z[g]$ (\ref{PartitionFunction})
instead of $Z[a]$ (\ref{Z_zeta_a2}), or at
least understanding better the mapping (\ref{eq:switching_to_imaginary_time})
since one is not guaranteed that $\rho_{\Lambda}$ is a constant
in general. This, however, would require the use of the full theory
of quantum gravity whose structure and mathematical apparatus are
not yet fully understood.
Nevertheless, some important results are already available
as pertained to a dynamical solution of the cosmological constant problem---Ref
\cite{Antoniadis&Mottola1992},
along with the possibility of numerical conformations using causal dynamical
triangulations \cite{Mottola1995,Ambjorn2004}.

The reasoning above shows that the quantum field theory estimate of
the vacuum energy density is probably valid only for sub-microscopic
(Planck size) universes; however, for macroscopic size systems, one
needs to use the $a$-scaling mapping to arrive at the observed value of $\Lambda$. 
An open question is the strict definition of the characteristic
scale and its meaning. It seems that there are many options: one can
think of the characteristic scale $\zeta$ as the size to which a
given universe has inflated during the inflation era, or the size
at which general relativity has become a good enough approximation
so one doesn't need quantum gravity, or the size at which $\Lambda$
is dominated by the zero point energy of the gravitational field and
all the matter contributions are negligible, or the size of the future
cosmological horizon, and so on and so on.
This issue, however, could be short circuited if one promotes the characteristic scale $a$
into a conformal gauge factor $\lambda(x)$ as part of the Integrable Weyl  Geometry framework.
For example, using FLRW metric:
\begin{equation}
ds^2=-c^2dt^2+R(t)^2\left(\frac{dr^2}{1-kr^2}+r^2d\Omega^2\right)=\lambda^2ds'^2
\label{FLRW}
\end{equation}

For non-zero Gaussian curvature $k$, one can consider a small enough patch 
of space $r<<1/\sqrt{|k|}$ to be Gaussian flat, that is, $k\approx 0$. 
Then, by switching to conformal time $\tau$ using $dt=R(t)d\tau$, one~can
identify conformal factor $\lambda=R(\tau)$ such that the new metric is 
conformally equivalent to the Minkowski metric and therefore a unimodular.
In the case of $k=0$, such interplay has been described explicitly in the 
discussion of the application to the empty space solution \cite{Maeder2017b}. 

Integrating out the matter fields is an essential step in the above calculations.
When applied in conjecture with the above reasoning, it leads to a
value of $\Lambda$, which is very close to the measured value perhaps
due to the fact that the value of $\Lambda$ calculated in this paper
is mainly due to the vacuum energy density of the classical gravitational
field in an $a$-derivable universe. It seems that the above reasoning
and results are consistent with the view that the effective cosmological
constant (\ref{eq:Lambda_eff}) is an integral of the motion with
a significant contribution from the Ricci scalar. Even more, it may be more beneficial 
and less puzzling if one is considering the smallness of $\Lambda$ to be due to
the large value of the entropy in the observed Universe.
Furthermore, if one is using the traceless version of the  Einstein's equations, then the 
zero point energy of the matter fields does not 
contribute to the vacuum energy density of the classical gravitational field. 
A unimodular gravity approach to Einstein's equations is well aligned 
with this view \cite{Smolin2009} and, if combined with the view that 
the cosmological constant is about the entropy of the system and not 
about its vacuum energy density as suggested by the BlackHole Cosmology 
(\ref{eq:SdS}), then one has a plausible resolution of the  main three
puzzles about the cosmological constant.

In summary, the key points of the paper are: the introduction of the
\textit{Multiverse Partition Function} for an ensemble of universes
enumerated by their \textit{characteristic size} $a$ that is also
used as integration variable in the partition function. The averaged
characteristic scale of the ensemble is estimated by using only members
that satisfy the Einstein field equations. The averaged characteristic
scale is compatible with the Planck length when considering an ensemble
built on Planck-scale seed universes with vacuum energy density of
order one; that is, $\tilde{\Lambda}\approx 8\pi$. For $a$-derivable
universes with a characteristic scale of the order of the observed
Universe $a\approx 6\times10^{60}$, the cosmological constant 
$\Lambda=\tilde{\Lambda}/a^{2}$
is in the range $10^{-121}\div10^{-122}$, which is close in magnitude
to the observed value $10^{-123}$.

The result that the overall averaged characteristic size of an ensemble of universes 
that satisfies the Einstein Field Equations (\ref{Einstein Equations}) and (\ref{EH Action for GR})
is of the order of the Planck scale is not obvious in general. 
Explaining the observed value of $\Lambda$ as due to the possibility that 
we live in a very large $a$-derivable universe might suggest that this is a low probability event. 
Thus, it may seem like an argument in support of the anthropic principle.
However, if the physical reality is such that the idea of a characteristic scale $\zeta$ 
is actually an illusion of our practical models, then the overall argument would not be applicable. 
For example, a scale-invariant paradigm based on the Weyl Integrable Geometry,
like the one discussed by the authors as an alternative to $\Lambda CDM$,
would make all homogeneous and isotropic $a$-derivable universes, 
based on a specific EFE homogeneous and isotropic solution $\{g_{\alpha\beta},\Lambda\}$, 
part of the same Universe via their time-dependent conformal factor
$\lambda(t)=\sqrt{\frac{3}{\Lambda}}\frac{1}{ct}$ \cite{Maeder2017b}. Such~an approach is 
clearly modifying significantly the main framework (\ref{Einstein Equations}) and
(\ref{EH Action for GR}), their consequent equivalent expressions, 
and of course the final expressions (\ref{Z_zeta_a2}) and (\ref{<zeta>})
that have lead to the result that the overall averaged characteristic size of an ensemble 
of universes is of the order of the Planck scale. 
Furthermore, utilizing the possibility of a conformal gauge factor  $\lambda(x)$ could lead to
unimodular gravity gauge choice where one considers the scale factor $a$ in 
$g_{\alpha\beta}=a^{2}\tilde{g}_{\alpha\beta}$ as an overall scaling 
field $a=\lambda(x)$ such that $\det{\tilde{g}_{\alpha\beta}}=1$.
Exploring the possible connection between unimodular gravity gauge 
and the Weyl Integrable Geometry 
by using the viewpoint on the conformal perturbations described by Mottola and his 
collaborations \cite{Mazur&Mottola1990,Antoniadis&Mottola1992} 
is an interesting research direction that  could bring new light to the Cosmological 
Constant problem, as well as on the Dark Matter problem via the SIV paradigm and 
probably a new understanding of the SIV conformal factor 
$\lambda(t)=\sqrt{\frac{3}{\Lambda}}\frac{1}{ct}$ within the framework of
quantum gravity.

\vspace{6pt} 
%%%%%%%%%%%%%%%%%%%%%%%%%%%%%%%%%%%%%%%%%%
%MDPI: Please add. 
%VGG added:
\authorcontributions{
V.G., is the lead author and researcher on the topics discussed by the paper.
He has analyzed the relevant literature and mathematical expressions.
A.M. has verified the mathematical correctness of the equations and the relevant physics conclusions. 
Both co-authors have been actively involved in the writing of the paper and its draft versions.}  

%%%%%%%%%%%%%%%%%%%%%%%%%%%%%%%%%%%%%%%%%%
\funding{This research received no external funding.}
%\funding{\hl{Please add:} ``This research received no external funding'' or ``This research was funded by NAME OF FUNDER Grant No.  XXX.'' and  and ``The APC was funded by XXX''. Check carefully that the details given are accurate and use the standard spelling of funding agency names at \url{https://search.crossref.org/funding}, any~errors may affect your future funding.}

%%%%%%%%%%%%%%%%%%%%%%%%%%%%%%%%%%%%%%%%%%
\acknowledgments{V.G. acknowledges useful discussions with 
%\hl{Dr.} 
%Titles (Dr., Mr., Prof., etc.) should NOT be used in the Acknowledgments section. 
%Please include first names here if possible
Carlos C. Perelman as well as discussions with 
Rob Macri, Emil M. Prodanov, and Tom Luu in 
the very early stage of his research on the topic. 
V. G. is extremely grateful to his wife and daughters for their understanding
and family support at various stages of the research presented. 
A. M. expresses his deep gratitude to his wife and to D. Gachet for their continuous support.}

%%%%%%%%%%%%%%%%%%%%%%%%%%%%%%%%%%%%%%%%%%
\conflictsofinterest{The authors declare no conflict of interest.} 
%\conflictsofinterest{\hl{Declare conflicts of interest} or state ``The authors declare no conflict of interest.''} 

%%%%%%%%%%%%%%%%%%%%%%%%%%%%%%%%%%%%%%%%%%

%\bibliographystyle{plane}
%\bibliographystyle{unsrt} 
%\bibliographystyle{apsrev}
%\bibliographystyle{aipnum4-1}
%\bibliographystyle{iopart-num}
%\bibliographystyle{phaip}

%\bibliographystyle{unsrt}\bibliography{CosmologicalConstant}\end{document}

\begin{thebibliography}{999}

\bibitem{Spergel2007}
Spergel, D.N.; Bean, R.; Doré, O.; Nolta, M.R.; Bennett, C.L.; Dunkley, J.; Hinshaw, G.; Jarosik, N.; Komatsu,~E.; Page, L.; et al.
\newblock {Three-Year Wilkinson Microwave Anisotropy Probe (WMAP) Observations:Implications for Cosmology}.
\newblock {\em  Astrophys. J. Suppl. Ser.} \textbf{2007}, \emph{170}, 377--408.

\bibitem{Komatsu2011}
Komatsu, E.; Smith, K.M.; Dunkley, J.; Bennett, C.L.; Gold, B.; Hinshaw, G.; Jarosik, N.; Larson, D.; Nolta, M.R.; Page, L.; et al.
\newblock {Seven-year Wilkinson Microwave Anisotropy Probe (WMAP) Observations: Cosmological Interpretation}.
\newblock {\em  Astrophys. J. Suppl. Ser.} \textbf{2011}, \emph{192}, 18.

\bibitem{PlanckCollaboration2016}
Ade, P.A.R.; Aghanim, N.; Arnaud, M.; Ashdown, M.; Aumont, J.; Baccigalupi, C.; Banday, A.J.; Barreiro,~R.B.; Bartlett, J.G.; Bartolo, N.; et al.
\newblock {Planck 2015 results. XIII. Cosmological parameters}.
\newblock {\em Astron. Astrophys.} \textbf{2016}, \emph{594}, A13.

\bibitem{Hawking1983}
{Hawking}, S.W.
\newblock {The cosmological constant}.
\newblock {\em Philos. Trans. R. Soc. Lond. Ser. A} \textbf{1983}, \emph{310}, 303--309.

\bibitem{Weinberg1989}
{Weinberg}, S.
\newblock {The cosmological constant problem}.
\newblock {\em Rev. Mod. Phys.} \textbf{1989}, \emph{61}, 1--23.

\bibitem{Coleman1988}
\textls[-15]{{Coleman}, S.
\newblock {Why there is nothing rather than something: A theory of the cosmological constant}.
\newblock {\em Nucl. Phys. B}} \textbf{1988}, \emph{310}, 643--668.

\bibitem{Hinshaw2013}
Hinshaw, G.; Larson, D.; Komatsu, E.; Spergel, D.N.; Bennett, C.L.; Dunkley, J.; Nolta, M.R.; Halpern, M.; Hill, R.S.; Odegard, N.; et al.
\newblock {Nine-year Wilkinson Microwave Anisotropy Probe (WMAP) Observations:
 Cosmological Parameter Results}.
\newblock {\em  Astrophys. J. Suppl. Ser.} \textbf{2013}, \emph{208}, 19.

\bibitem{Valentino2020}%{2020NatAs...4..196D} 
Di Valentino, E.; Melchiorri, A.; Silk, J. 
Planck evidence for a closed Universe and a possible crisis for cosmology. 
\emph{Nat. Astron.} \textbf{2020}, \emph{4}, 196.

\bibitem{Carroll1992}
{Carroll}, S.M.; {Press}, W.H.; {Turner}, E.L.
\newblock {The cosmological constant}.
\newblock {\em Annu. Rev. Astron Astrophys.} \textbf{1992}, \emph{30}, 499--542.

\bibitem{Smolin2009}%{2009PhRvD..80h4003S} 
Smolin, L. 
\newblock {Quantization of unimodular gravity and the cosmological constant problems}.
\newblock {\em Phys. Rev. D} \textbf{2009}, \emph{80}, 084003.

\bibitem{Farag&Ali&Das2015}%{2015PhLB..741..276F} 
Ali, A.F.; Das, S.
\newblock {Cosmology from quantum potential.}
\newblock {\em Phys. Lett. B} \textbf{2015}, \emph{741}, 276.

\bibitem{Chiarelli2019}%{2017arXiv171106093C} 
Chiarelli, P.
\newblock {The Gravity of the Classical Klein-Gordon Field.}
\newblock {\em Symmetry} \textbf{2019}, \emph{11}, 322.
%[arXiv:1711.06093]

\bibitem{Chiarelli2020}%{2017arXiv171106545C} 
Chiarelli, P.
\newblock {The Spinor-Tensor Gravity of the Classical Dirac Field.}
\newblock {\em Symmetry} \textbf{2020}, \emph{12}, 1124.
%[arXiv:1711.06545]

\bibitem{Antoniadis&Mottola1992}%{1992PhRvD..45.2013A} 
Antoniadis, I.; Mottola, E.
\newblock {Four-dimensional quantum gravity in the conformal sector.}
\newblock {\em Phys. Rev. D} \textbf{1992}, \emph{45},~2013.

\bibitem{Mottola1995}%{1995JMP....36.2470M} 
Mottola, E. 
\newblock {Functional integration over geometries.}
\newblock {\em J. Math. Phys.} \textbf{1995}, \emph{36}, 2470.
%[arXiv:hep-th/9502109]

\bibitem{Ambjorn2004}
{Ambj{\o}rn}, J.; {Jurkiewicz}, J.; {Loll}, R.
\newblock {Emergence of a 4D World from Causal Quantum Gravity}.
\newblock {\em Phys. Rev. Lett.} \textbf{2004}, \emph{93}, 131301.

\bibitem{Nottale2008}
{Nottale}, L.
\newblock {Scale relativity and fractal space-time: Theory and applications}.
\newblock {\em Found. Sci.} \textbf{2010}, \emph{15}, 101--152.

\bibitem{Lucat2018}
{Lucat}, S.; {Prokopec}, T.; {Swiezewska}, B.
\newblock {Conformal symmetry and the cosmological constant problem}.
\newblock {\em Int. J. Mod. Phys. D} \textbf{2018}, \emph{27}, 1847014.

\bibitem{Pathria1972}
{Pathria}, R.K.
\newblock {The Universe as a Black Hole}.
\newblock {\em Nature} \textbf{1972}, \emph{240}, 298--299.

\bibitem{Perelman2018a}
Perelman, C.C.
\newblock Is Dark Matter and Black-Hole Cosmology an Effect of Born's Reciprocal Relativity Theory?
\newblock {\em Can. J. Phys.} \textbf{2019}, \emph{97}, 198--209.

\bibitem{Brans1961}
{Brans}, C.; {Dicke}, R.H.
\newblock {Mach's Principle and a Relativistic Theory of Gravitation}.
\newblock {\em Phys. Rev.} \textbf{1961}, \emph{124}, 925--935.

\bibitem{Li2011}
{Li}, M.; {Li}, X.-D.; {Wang}, S.; {Wang}, Y.
\newblock {Dark Energy}.
\newblock {\em Commun. Theor. Phys.} \textbf{2011}, \emph{56}, 525--604.

\bibitem{Poplawski2018}%{2016ApJ...832...96P} 
{Pop{\l}awski}, N.
\newblock {Universe in a Black Hole in Einstein-Cartan Gravity}
\newblock {\em  Astrophys.~J.} \textbf{2017}, \emph{832}, 158.
%MDPI: Please add.

\bibitem{Gibbons&Hawking1977}%{1977PhRvD..15.2752G} 
Gibbons, G.W.; Hawking, S.W.
\newblock {Action integrals and partition functions in quantum gravity}. 
\newblock {\em Phys. Rev. D} \textbf{1977}, {\emph{15}}, 2752.


\bibitem{deSitter1917c}%{1917MNRAS..78....3D} 
De Sitter, W.
\newblock {Einstein's theory of gravitation and its astronomical consequences. Third paper}.
\newblock {\em Mon. Not. R. Astron. Soc.} \textbf{1917}, \emph{78}, 3.

\bibitem{Maeder2017b}
{Maeder}, A.
\newblock {An Alternative to the {$\Lambda$}CDM Model: The Case of Scale Invariance}.
\newblock {\em  Astrophys. J. Suppl. Ser.} \textbf{2017}, \emph{834}, 194.

\bibitem{Maeder2017}
{Maeder}, A.
\newblock {Dynamical Effects of the Scale Invariance of the Empty Space: The Fall of Dark Matter?}
\newblock {\em  Astrophys.~J.} \textbf{2017}, \emph{849}, 158.

\bibitem{Mazur&Mottola1990}%{1990NuPhB.341..187M} 
Mazur, P.O.; Mottola, E.
\newblock {The path integral measure, conformal factor problem and stability of the ground state of quantum gravity.}
\newblock {\em Nucl. Phys. B} \textbf{1990}, \emph{341}, 187.

\bibitem{MaedGueor19} 
Maeder, A.; Gueorguiev, V.G. 
\emph{Phys. Dark Universe} \textbf{2019}, \emph{25}, 100315.

\bibitem{Vilenkin2001}%{2001hep.th....6083V} 
Vilenkin, A.
\newblock {Cosmological constant problems and their solutions.}
\newblock {\em Astron. Cosmol. Fundam. Phys.} \textbf{2003}, 70--78. 
doi:10.1007/10857580\_7.
%Newly added information, please confirm. %VGG yes it is correct.
%[arXiv e-prints hep-th/0106083].

\bibitem{Kawai&Okada2011}%{2011IJMPA..26.3107K} 
\textls[-15]{Kawai, H.; Okada, T.
\newblock {Asymptotically Vanishing Cosmological Constant in the Multiverse.}
\newblock {\em Int. J. Mod. Phys. A}} \textbf{2011}, \emph{26}, 3107.


\bibitem{Pejhan2018}
{Pejhan}, H.; {Bamba}, K.; {Enayati}, M.; {Rahbardehghan}, S.
\newblock {A small non-vanishing cosmological constant from the
 Krein-Gupta-Bleuler vacuum}.
\newblock {\em Phys. Lett. B} \textbf{2018}, \emph{785}, 567--569.

\bibitem{Sahni2000}
{Sahni}, V.; {Starobinsky}, A.
\newblock {The Case for a Positive Cosmological {$\Lambda$}-Term}.
\newblock {\em Int. J. Mod. Phys. D} \textbf{2000}, \emph{9}, 373--443.

\bibitem{Hawking1984b}%{1984PhLB..134..403H} 
Hawking, S.W.
\newblock {The cosmological constant is probably zero.}
\newblock {\em Phys. Lett. B} \textbf{1984}, \emph{134}, 403.

\bibitem{Vilenkin2020}%{2020PhRvD.101d3520V} %hep-th-1912.02187
Vilenkin, A.; Yamada, M.
\newblock {Four-volume cutoff measure of the multiverse.} 
\newblock {\em Phys. Rev. D} \textbf{2020}, {\emph{101}}, 043520.

\bibitem{Vilenkin1984}%{1984PhRvD..30..509V} 
Vilenkin, A.
\newblock {Quantum creation of universes.}
\newblock {\em Phys. Rev.  D} \textbf{1984}, \emph{30}, 509.








\end{thebibliography}

\reftitle{References}

\end{document}